\documentclass[twocolumn,showpacs]{revtex4}

\usepackage{graphicx}
\usepackage{dcolumn}
\usepackage{amsmath}
\usepackage{longtable}
\begin{document}

\title{Fermi surface of superconducting LaFePO determined from quantum oscillations}
\author{A.I. Coldea$^1$, J.D. Fletcher$^1$, A. Carrington$^1$, J.G. Analytis$^2$, A.F. Bangura$^1$,
J.-H. Chu$^2$, A. S. Erickson$^2$, I.R. Fisher$^2$, N.E. Hussey$^1$,
and R.D. McDonald$^3$}

\affiliation{$^1$H.H. Wills Physics Laboratory, University of
Bristol, Tyndall Avenue, Bristol, United Kingdom.}

\affiliation{$^2$Geballe Laboratory for Advanced Materials and
Department of Applied Physics, Stanford University, Stanford,
California 94305-4045}

\affiliation{$^3$National High Magnetic Field Laboratory, Los Alamos
National Laboratory, MS E536, Los Alamos, New Mexico 87545, USA}

\begin{abstract}
We report extensive measurements of quantum oscillations in the
normal state of the Fe-based superconductor LaFePO, ($T_c \sim 6$~K)
using low temperature torque magnetometry and transport in high
static magnetic fields (45~T). We find that the Fermi surface
is in broad agreement with the band-structure calculations
with the quasiparticle mass enhanced by a factor $\sim$2. The
quasi-two dimensional Fermi surface consist of nearly-nested
electron and hole pockets, suggesting proximity to a spin/charge
density wave instability.

\end{abstract}


\date{\today}
\maketitle

The recent discovery of superconductivity in ferro-oxypnictides
\cite{Kamihara2006,Ren2008}, has generated huge interest as
 as another possible route towards achieving high $T_c$
superconductivity. LaFePO was among the first Fe-based
superconductor to be discovered and has a transition temperature of
up to $T_{c} \approx 7$~K \cite{Kamihara2006}. This compound is
isostructural with LaFeAsO, which is non-superconducting and has a
spin-density wave (SDW) ground state \cite{delaCruz}, but becomes a
relatively high-$T_c$ superconductor ($T_{c} \approx 25$~K) when
electron doped \cite{Kamihara2008}. By changing the rare-earth ion,
$T_c$ reaches 55~K in SmFeAsO$_{1-x}$F$_x$
[Ref.~\onlinecite{Ren2008}]. Theoretical models suggest that the
pairing mechanism in the Fe-based superconductors may be mediated by
magnetic fluctuations due to the proximity to a SDW
\cite{Mazin2008SC,Chubukov2008,Cvetkovic2008}. Knowing the fine
details of the Fermi surface topology, its tendency towards
instabilities as well as the strength of the coupling of the
quasiparticles to excitations is important for understanding the
superconductivity.

 Quantum oscillations provide a bulk probe
 of the
electronic structure, giving detailed information about the Fermi
surface (FS) topology and mass renormalization. To observe quantum
oscillations samples must be extremely clean and the upper critical
field must be low enough for the normal state to be accessed; LaFePO
is a material which fulfils both these requirements. The tetragonal
layered structure of LaFePO is made of alternating highly conductive
FeP layers and poorly conducting LaO layers stacked along the $c$
axis \cite{Kamihara2006}, hence the Fermi surface is expected to be
quasi-two dimensional \cite{Lebègue2007}. Here we report extensive
measurements of quantum oscillations in torque and transport data.
We find that the Fermi surface of LaFePO is composed of quasi
two-dimensional nearly-nested electron and hole pockets with
moderate enhancement of the quasiparticle masses.

  Single crystals of LaFePO, with dimensions up
to $0.2 \times 0.2 \times 0.04$ mm$^{3}$, and residual resistance
ratios $\rho$(300~K)/$\rho$(10~K) up to 85, were grown from a tin
flux \cite{Analytis2008}. Single crystal x-ray diffraction gives
lattice parameters $a = 3.941(2)$~\AA, $c =8.507(5)$~\AA, and La/P
positions $z_{\rm La}=0.148901(19)$, $z_{\rm P}=0.63477(10)$ in
agreement with previous results \cite{Kamihara2006}.
Torque measurements were performed with piezoresistive
microcantilevers \cite{cantilever} down to 0.3~K on different single
crystals from the same batch ($T_c \approx 6$~K); one in Bristol  up
to 18~T (sample B) and another crystal at the NHMFL, Tallahassee, up
to 45~T (sample A). Interplane electrical transport has been
measured on a third sample (sample C).

\begin{figure}
\centering
\includegraphics[width=8cm]{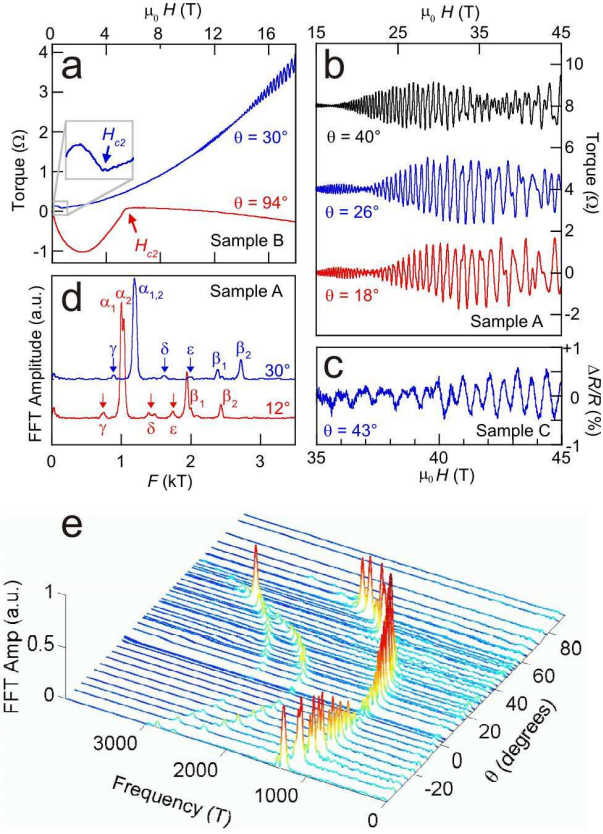}
\caption{(color online) a) Torque measurements on LaFePO measured at
$T = 0.35$~K and in magnetic fields up to 18~T for different
magnetic field directions. The arrows indicate the position of
$H_{\rm c2}$. b) Oscillatory part of the torque (dHvA) and c)
resistance (SdH) in magnetic fields up to 45~T. d) Fourier transform
spectra (field range 15-45~T) for two different magnetic field
directions. e) The angle dependence of the Fourier transform spectra
for the field range 10-18~T (sample B).} \label{figure1}
\end{figure}
Figure 1a shows raw torque measurements versus magnetic field at low
temperatures ($T = 0.3$~K) at various angles $\theta$ between the
magnetic field direction and the $c$ axis \cite{angle}. At low
fields we observe behavior typical of a bulk anisotropic type-II
superconductor in the vortex state \cite{Hao1991}. The signal is
reversible, indicating weak pinning of vortices. The upper critical
field (Fig. 1a) is strongly anisotropic, varying between
$\mu_0H_{\rm c2}= 7.2$~T when the magnetic field is parallel to the
{\it ab} plane and $\mu_0H_{\rm c2} =$ 0.68~T when $B||c$. For
magnetic fields above $\sim 9$~T, we observe oscillations periodic
in inverse field, which arise from the de Haas-van Alphen (dHvA)
effect. The oscillatory signal is clearly visible by subtracting a
monotonic background (a fifth-order polynomial) from the torque data
(dHvA in Fig.1b) or transport data [Shubnikov-de Haas effect (SdH)
in Fig.1c].

From the fast Fourier transform (FFT) spectra of the oscillatory
data (Figs. 1d and 1e) we can identify the dHvA frequencies $F$,
which are related to the extremal cross-sectional areas $A_{k}$, of
the FS orbits via the Onsager relation, $F = (\hbar/2\pi e) A_{k}$.
From the evolution of these frequencies as the magnetic field is
rotated from being along the $c$ axis ($\theta = 0^{\circ}$) towards
the $a$ axis ($\theta = 90^{\circ}$) \cite{angle}, we can construct
a detailed three dimensional picture of the shape and size of the
Fermi surface. Four frequencies have significantly larger amplitudes
than the others in both samples A and B as shown in Figs. 1d and 2a.
We label $\alpha_1$ and $\alpha_2$ the two closely split frequencies
at $F \approx 1$~kT ($\Delta F \approx 35$~T) and the two higher
frequencies, $\beta_1$ and $\beta_2$. Besides these intense features
(and their harmonics \cite{cantilever}), in sample A (which was
measured to much higher fields) we see several other frequencies
with smaller amplitudes; these are labeled $\gamma$, $\delta$,
$\varepsilon$ and have frequencies in the range 0.7--1.7~kT. An
extremely weak signal, $\eta$, was observed only in sample B
(measured up to 18~T) which we believe originates from a small
misaligned crystallite \cite{angle}. The observed frequencies
correspond to a fraction varying between 2.8\% to 9\% of the basal
plane area of the Brillouin zone. They are significantly larger than
those observed in the double-layer Fe-As compound, SrFe$_2$As$_{2}$
\cite{Sebastian2008}, which is non-superconducting, and is likely to
have a reconstructed Fermi surface at low temperatures due to a
spin-density-wave ground state.

Figs.~1e and 2a show the angular dependence of the main frequencies
and their amplitudes. For a purely two-dimensional cylindrical Fermi
surface the dHvA frequency varies like $1/\cos{\theta}$ and
deviations from this indicate the degree of warping for a
quasi-two-dimensional cylinder.
 As shown in Fig.~2b, the orbits $\beta_1$ and
$\beta_2$ both originate from sections of FS which have significant
warping, but with opposite curvature (i.e., maximal and minimal
areas respectively). The angle dependence of these orbits is well
described by Yamaji's analysis of simple cosine warping of a
two-dimensional cylinder \cite{Yamajii1989}, as is the large
increase in the amplitude of the oscillations at $\theta \sim
45^{\circ}$, where the frequencies cross (see Fig. 1e). These
observations indicate that $\beta_1$ and $\beta_2$ arise from the
same FS sheet. For the $\alpha$ orbits the warping of the FS sheet
is very small, but the almost identical amplitudes, effective masses
(see below) and frequencies strongly suggests that these orbits
arise from a single, weakly warped, FS sheet.

\begin{figure}
\centering
\includegraphics[width=8cm]{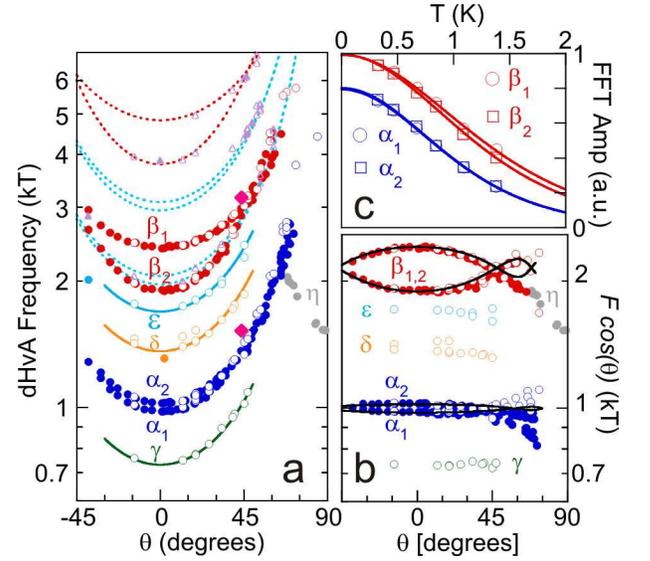}
\caption{(color online) a) Angle dependence of all observed
frequencies. Different symbols correspond to sample A (open
circles), sample B (filled circles) and sample C (diamonds).
Possible harmonics of the main frequencies are shown by triangles
and the dashed lines indicate their calculated location. Solid lines
are fits to a 1/cos $\theta$ dependence. b) Angle dependence of $F
\cos$ $\theta$. Solid lines are calculations for a simple cosine
warped cylinder. c) The temperature dependence of the Fourier
amplitude for $\theta = 32^\circ$ (data are offset for clarity).
Solid lines are fits to the Lifshitz-Kosevich formula
\cite{Shoenberg1984}. The obtained values of the effective mass are
listed in Table I. }\label{figure2}\end{figure}

The effective mass of the quasiparticles on the various orbits were
determined by fitting the temperature dependent amplitude of the
oscillations to the conventional Lifshitz-Kosevich
formula\cite{Shoenberg1984}, as shown in Fig.~2c. The obtained
masses range between 1.7~{\it m$_e$} and 2.1{\it m$_e$} and are
listed in Table~I ($m_e$ is the free electron mass).
\begin{table}
\centering \caption{Effective masses ($m^*$) and frequencies from
dHvA data for samples A and B. Calculated band masses ($m_b$) in
LaFePO for shifted and unshifted bands. Orbits are labeled by band
number and the location of their center.}
\begin{tabular}{c  c c c | c c c}
\hline \hline
Branch      &  $F$(kT)      & \multicolumn{2}{c}{$m^*/m_e$}  & Orbit    &  \multicolumn{2}{c}{$m_b/m_e$}\\
(exp.)      &               & Sample A &  Sample B & (calc.)  & unshifted & shifted\\
    \hline
$\alpha_1$  &   0.985(7)    &   1.9(2)  &  1.8(1) & 1Z&1.0 & 1.9\\
$\alpha_2$  &   1.025(7)    &   1.9(2)  &  1.8(1) & 2$\Gamma$&0.9 &0.7\\
$\beta_1$   &   1.91(1)     &   1.7(2)  &  1.8(1) & 2Z&1.2 &1.1\\
$\beta_2$   &   2.41(1)     &   1.8(2)  &  1.9(1) & 3$\Gamma$&1.8 & 1.1\\
$\delta$    &   1.36(2)     &   1.8(3)  &         & 3Z&2.9 & 2.5\\
$\gamma$    &   0.73(2)     &   1.7(3)  &         & 4M&0.7 & 0.7\\
$\epsilon$  &   1.69(2)     &   2.1(3)  &         & 4A&0.9 & 0.9\\
            &               &           &         & 5M&0.8 & 0.7\\
            &               &           &         & 5A&0.9 & 0.8\\
\hline \hline
\end{tabular}
\end{table}

\begin{figure}
\centering
\includegraphics[width=8cm]{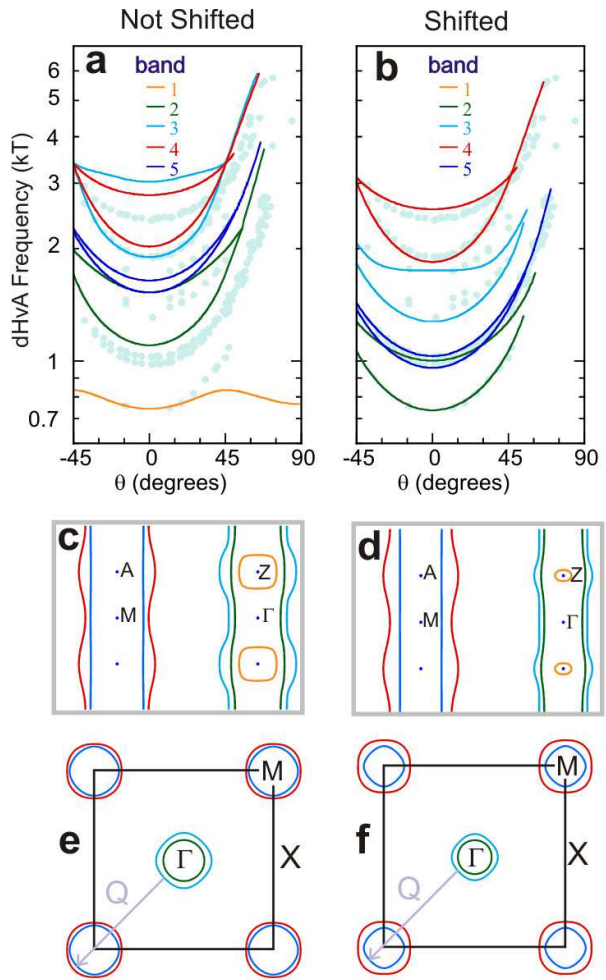}
\caption{(color online) a) Calculated dHvA frequencies versus angle
for unshifted bands compared to experimental data (shown in light
blue) b) As in a) but for shifted bands (see text). c), d)
Two-dimensional (110) cuts of the Fermi surface and e), f)
two-dimensional cuts in the $\Gamma$XM plane (001) for unshifted and
shifted bands, respectively. Solid black lines indicate the
Brillouin zone. A possible inter-band nesting vector, {\bf
Q}$~\approx$~[$\pi, \pi, 0$], between electron and hole bands is
shown.} \label{figure3}
\end{figure}

We now compare our experimental observations with predictions of the
density functional theory calculations of the electronic structure
of LaFePO. Our calculations were made using the WIEN2K code with the
experimental lattice parameters and atomic positions and including
spin-orbit interactions \cite{Blaha2001}. The resulting band
structure is in good agreement with that reported by Leb\'egue
\cite{Lebègue2007}. The density of states at the Fermi level are
derived mainly from Fe and P bands suggesting that the carriers flow
mainly in the 2D FeP layer. The Fermi surface mainly comprises
small, slightly warped tubular sections running along the $c$
direction. There are two hole cylinders centered on the Brillouin
zone centre ($\Gamma$) and two electron cylinders centered on the
zone corner (M) (see Fig. 3c-f). In addition, there is a small
three-dimensional (3D) hole pocket centered on Z. The spin-orbit
interaction makes small but significant changes to the Fermi
surfaces; most notably it breaks the degeneracy of the bands
crossing the Fermi level along the XM line such that the two
cylindrical Fermi surfaces no longer touch and it also increases the
separation of the elliptical hole pocket and the two cylindrical
hole surfaces (Fig. 3c).

The frequencies of the extremal dHvA orbits obtained from the
calculated band structure are compared with the experimental data in
Fig. 3a. The calculation predicts that there should be 9 frequencies
(two for each tube plus one for the 3D hole pocket) in the range
$0.7-3$~kT (for $\theta \simeq 0^\circ$), which is broadly similar
to what is observed experimentally. In particular orbits $\beta_1$
and $\beta_2$ closely resemble those expected from the larger
electron cylinder in both frequency and curvature. The shape and
splitting of orbits $\alpha_1$ and $\alpha_2$ are similar
to that of the smaller electron cylinder.
The larger amplitude of these oscillations indicates that they both
suffer significantly less damping than the other orbits (we estimate
a mean free path of $\approx 1300$~\AA~ and 800~\AA~ for $\alpha$
and $\beta$ orbits, respectively). This may be a natural consequence
of the fact that both of these electron orbits originate from the
same piece of Fermi surface in the larger unfolded Brillouin zone
(corresponding to the Fe-sublattice \cite{Mazin2008SC}). Assuming
the above assignment for the $\alpha$ and $\beta$ frequencies, the
three remaining frequencies ($\gamma$, $\delta$, $\varepsilon$)
would then correspond to hole orbits, although their exact
assignment is less clear. For these hole orbits the scattering is a
factor $\sim 2$ larger than the electron orbits (Figs. 1d and 1e).

Performing small rigid shifts of the energies of the electron and
hole bands improves the agreement with experiment. The two bands
giving rise to the electron orbits $\alpha$ and $\beta$ are shifted
by $-83$~meV and $-30$~meV, respectively and the hole bands by
$+53$~meV (Fig. 3b).
The band that gives rise to the 3D pocket (which we do not observe
experimentally) influences significantly the degree of warping along
the $c$-axis of the hole cylinders; if this band was absent we would
have a better agreement between our data and calculations (Figs. 3a
and 3b).

A two-dimensional cut in the $\Gamma$XM plane for our calculated
unshifted and shifted Fermi surface (see Figs.~3e and 3f) shows that
the electron pockets at the corner of the Brillouin zone (M) have
almost similar shapes and sizes to the hole pockets at the centre of
the zone ($\Gamma$). Nesting requires a perfect match between the
size and the Fermi surface topology  of the electron and hole
pockets and
this could stabilize a spin density wave (magnetic order) or charge
density wave (structural order) with a wave vector {\bf
Q}$~\approx$~[$\pi, \pi, 0$]
\cite{Mazin2008,Chubukov2008,Cvetkovic2008}.
LaFePO is non-magnetic \cite{McQueen2008} but importantly our
results suggest that size and the shape of the electron and hole
pockets are indeed very similar (Figs. 3e and f) implying that
LaFePO may be close to nesting and hence to a spin/charge density
wave instability.

For an undoped LaFePO the volume of the electron and hole sheets
should be equal (compensated metal) but shifting the bands (as
described above) creates an imbalance of $\sim0.03$ electrons per formula unit.
A possible explanation for this imbalance can be related to a small
amount of electron doping in our crystals due to oxygen
non-stoichiometry ($\sim 1.5\%$ oxygen deficiency which is below the
resolution of our x-ray data).

The band structure calculation allows us to estimate the many-body
(electron-phonon and electron-electron) enhancements of the
quasiparticle masses over their band values. Table I shows the band
masses for all the predicted orbits as calculated, and after
application of band shifts. For the identified bands a moderate
renormalisation is found, with $m^*/m_b =$ ($1+\lambda$) $\approx
2$. For the hole orbits the calculated frequencies do not correspond
exactly to the experimental ones, however, as the band masses for
most of the hole orbits are in the range $0.7-1.1~ m_e$ and the
measured $m^*$ values are $\approx 2 m_e$, a similar level of
enhancement is likely. Reported measurements of the electronic
specific heat coefficient ($\gamma$) vary between 7-12 mJ/mol K$^{2}
~$\cite{Analytis2008,Kohama2008,McQueen2008}. In 2D materials
$\gamma$ can be estimated from the renormalised dHvA masses, summed
over all of the observed FS sheets. Four quasi-2D FS sheets with an
effective mass $\sim 2 m_e$ would give $\gamma_{\rm dHvA}\sim
6$~mJ/mol K$^{2}$, close to the lower end of the experimentally
observed values, without including any contribution from a 3D
pocket.

A recent ARPES study of LaFePO \cite{Lu2008} showed an electron-like
FS pocket centered at the M point and two hole sheets at the
$\Gamma$ point. The electron sheet and the smaller hole orbit have
areas close to the experimentally observed dHvA frequencies. When
compared to band structure calculations, the bands found in the
ARPES measurements are renormalized by a factor two, similar to the
mass enhancements we have found. This suggests that the
renormalization occurs over the whole band, rather than being
localized to energies close to the Fermi level, as found in
Sr$_2$RuO$_4$, \cite{Mackenzie1996}. One of the hole pockets seen in
ARPES is much larger ($>12$~kT) than any dHvA frequencies observed
here; its presence creates a significant charge imbalance of one
electron per unit cell, suggesting that this feature is related to
surface effects \cite{Lu2008}.

As in the other ferrooxypnictides, in LaFePO the bandstructure is
very sensitive to the position of the P atom
\cite{Vildosola2008,Mazin2008}. Considering the P atom in the
position optimized (the `relaxed' structure) results in hole orbits
which are much too {\it small} to explain our data. Much larger band
energy shifts are needed to bring it into approximate agreement and
we find that the band structure calculated with the experimental P
position provides a better description of the data.

In conclusion, we have found experimentally that the Fermi surface
of the superconductor LaFePO, is in broad agreement with band
structure calculations with a mass enhancement of about 2 due to
many-body interactions. The difference in amplitudes of the dHvA
signal between the hole and electron pockets is an indication of
different scattering rates affecting these orbits. The near-perfect
matching between the hole and the electron orbits that we observe,
suggests that LaFePO may be close to a spin/charge density wave
transition and that magnetic fluctuations are an important
ingredient in the physics of the Fe-based superconductors.

We thank E.A. Yelland, N. Fox, MF Haddow for technical help
and I. Mazin for helpful comments. This work was supported
financially by EPSRC (UK) and the Royal Society. AIC is grateful to
the Royal Society for a Dorothy Hodgkin Fellowship. Work at Stanford
was supported by the DOE, Office of Basic Energy Sciences under
contract DE-AC02-76SF00515. Work performed at the NHMFL in
Tallahassee, Florida, was supported by NSF Cooperative Agreement No.
DMR-0654118, by the State of Florida, and by the DOE.


\begin{thebibliography}{Bibliography}
\bibitem{cantilever}
Changes in lever resistance were measured with a Wheatstone bridge
circuit. The sensitivity of the levers is approximately 0.1$^\circ$
per Ohm. At certain angles and at the highest fields, the deflection
of the cantilever
is large enough to generate harmonics of the main dHvA frequencies.
\bibitem{angle}
The X-ray laue diffraction indicates that the exact axis of rotation
between is $\sim 35^\circ$ (sample A) and $\sim 14^\circ$ (sample B)
from the $a$ axis; a small piece of crystal
with its $c$-axis oriented 73$^\circ$ away from that of the main
piece of the crystal was detected in sample B.
\end{thebibliography}
\end{document}